\documentclass[10pt,journal,compsoc]{IEEEtran}
\ifCLASSOPTIONcompsoc
  \usepackage[nocompress]{cite}
\else
  \usepackage{cite}
\fi
\ifCLASSINFOpdf
   \usepackage[pdftex]{graphicx}
\else
   \usepackage[dvips]{graphicx}
\fi
\usepackage{subfigure}

\makeatletter
\newcommand{\removelatexerror}{\let\@latex@error\@gobble}
\makeatother

\usepackage{color}
\usepackage{xcolor}
\usepackage{soul}
\usepackage{multirow}
\usepackage{array}

\newcommand{\sysName}{\textit{SmartSmudge}}

\begin{document}
%
\title{{Region-aware Color Smudging}}
%
%
%
%

\author{Ying Jiang, Pengfei Xu, Congyi Zhang, Hongbo Fu, Henry Lau, Wenping Wang
\IEEEcompsocitemizethanks{
\IEEEcompsocthanksitem 
Y. Jiang (yingjiang@connect.hku.hk),H. Lau (henryyklau@gmail.com) are  with the University of Hong Kong. W. Wang (wenping@cs.hku.hk ) is with the University of Hong Kong and Texas A\&M University.
\IEEEcompsocthanksitem  C. Zhang (zhcongyi@gmail.com) is with the University of British Columbia. 
\IEEEcompsocthanksitem P. Xu (xupengfei.cg@gmail.com) is with Shenzhen University. 
\IEEEcompsocthanksitem H. Fu (hongbofu@cityu.edu.hk) is with  the City University of Hong Kong. 

}
}

%
%

\markboth{
IEEE Transactions on Visualization and Computer Graphics
}%
{Shell \MakeLowercase{\textit{et al.}}: Bare Demo of IEEEtran.cls for Computer Society Journals}
%



\IEEEtitleabstractindextext{%
\begin{abstract}
Color smudge operations from digital painting software enable users to create natural shading effects in high-fidelity paintings by interactively mixing colors. To precisely control results in traditional painting software, users tend to organize flat-filled color regions in multiple layers and smudge them to generate different color gradients. However, the requirement to carefully deal with regions makes the smudging process time-consuming and laborious, especially for non-professional users. This motivates us to investigate how to infer user-desired smudging effects when users smudge over regions in a single layer. To investigate improving color smudge performance, we first conduct a formative study. Following the findings of this study, we design~\sysName, a novel smudge tool that offers users dynamical smudge brushes and real-time region selection for easily generating natural and efficient shading effects. We demonstrate the efficiency and effectiveness of the proposed tool via a user study and quantitative analysis. 
\end{abstract}

}

\maketitle

\IEEEdisplaynontitleabstractindextext

%
\IEEEpeerreviewmaketitle

\IEEEraisesectionheading{\section{Introduction}\label{sec:introduction}}

%
%
%
%

\begin{figure}[t]
\centering
  \includegraphics[width=0.82\linewidth]{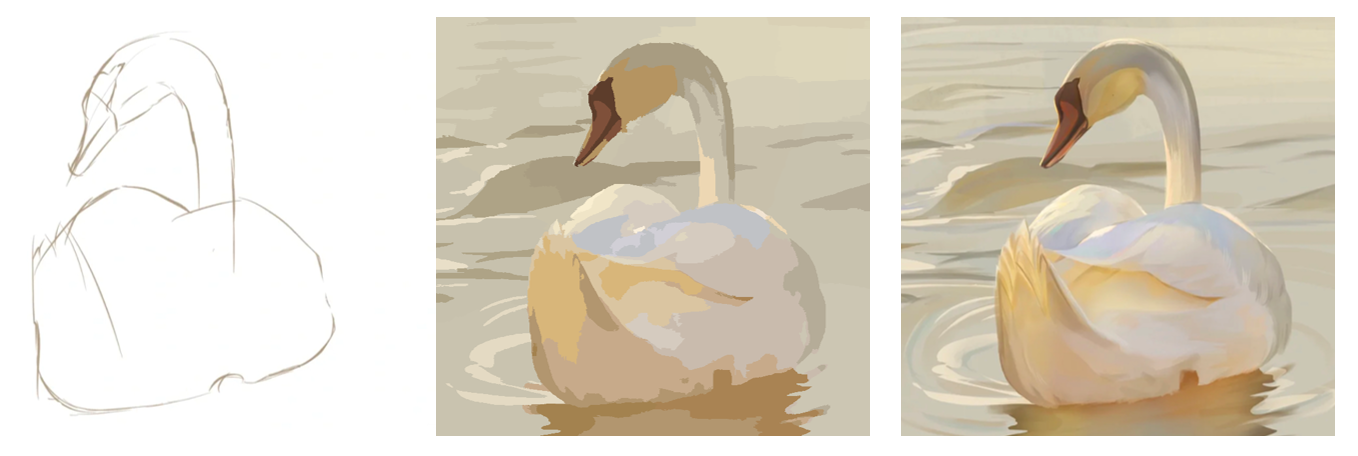}
\caption{Left: a sketch drawn by a user. Middle: a flat-filled painting created by filling colors in the sketch. Right: the final painting created by blurring and smudging the flat-filled painting.}
\label{fig:stage}
\end{figure}

\IEEEPARstart{D}{igital} painting is extensively used in graphic design and industrial fields, such as digital games, movie, and anime industries. While many ordinary users are interested in painting, creating high-fidelity paintings is often difficult for novices or amateur users and time-consuming for professional artists. Typically, a digital painting is created from scratch in three steps (Fig. \ref{fig:stage}): first, make a sketch outlining a desired shape; second, choose a proper color scheme and create a flat-filled painting; third, create the final painting by blending the colors in the flat-filled painting with smudge or blur tools. Among these three steps, the last one is the most challenging one~\cite{annum2014digital}. While various drawing assistant tools \cite{lee2011shadowdraw, dixon2010icandraw, lu2013realbrush, williford2019drawmyphoto} have been proposed to help even novice users make desired sketches in the first step, how to facilitate the creation of gradient-like shading effects is lack of concern. Existing sketch colorization techniques are able to automatically colorize sketches given user-specified color strokes, color dots, text descriptions \cite{zou2019language, zhang2017real, xiao2019interactive, qu2006manga, zhang2018two} or reference images \cite{akita2020colorization}. However, they are more suitable for creating paintings with flat filling or cartoon/manga shading styles that lack soft blending and detailed control of complex shading effects.

\begin{figure}[t]
\centering
  \includegraphics[width=0.8\linewidth]{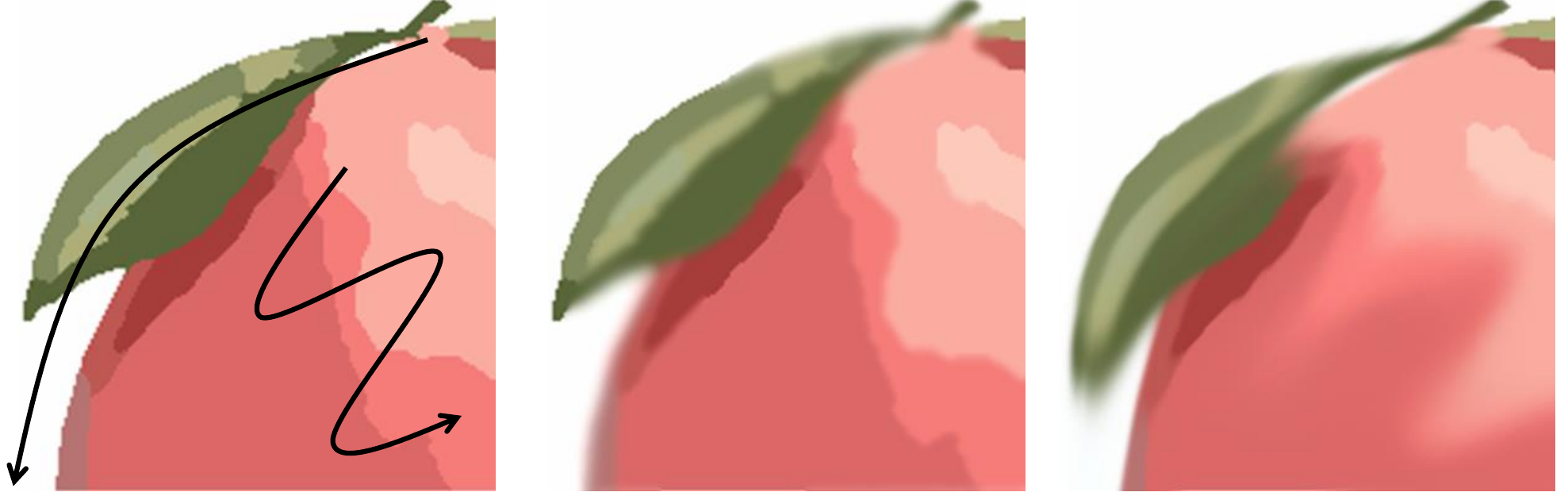}
\caption{Left: the original painting and smudge strokes. Middle and Right: the respective shading effects after a blur operation and a smudge operation with the same strokes and the brush size is in 50 pixels.}
\label{fig:blurvssmudge}
\end{figure}

\begin{figure}[t]
\centering
  \includegraphics[width=0.8\linewidth]{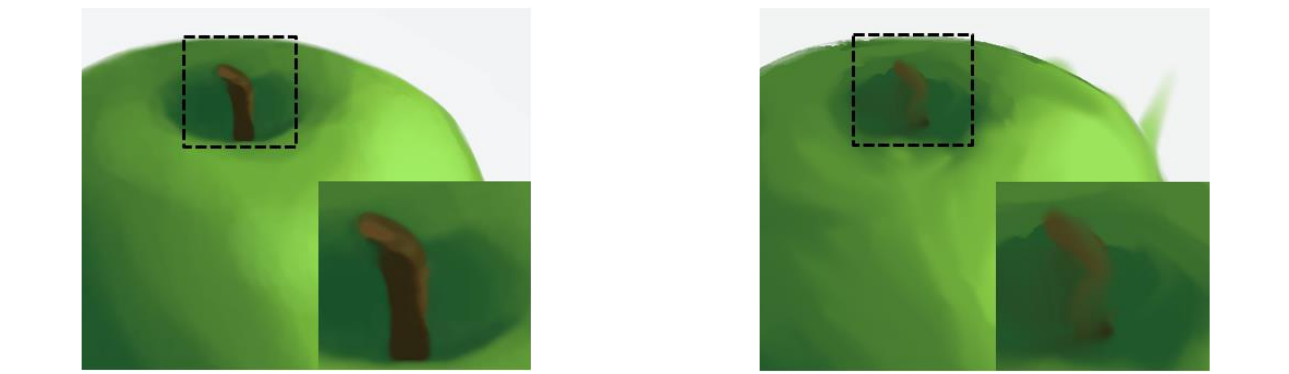}
\caption{{Left: natural shading effects by our proposed tool \emph{SmartSmudge}. Right: undesired shading effects caused by the traditional smudge tool.} 
}
\label{fig:prepost}
\end{figure}

In a digital painting process, users often repeatedly \cite{shugrina2017playful}  {by using the smudge or blur tool \cite{photoshop, clipstudio, procreate}. Different from the blur tool, which only blurs edges between different color regions, the smudge tool not only blurs edges but also liquefies colors along with the directions of smudge strokes, as illustrated in Fig. \ref{fig:blurvssmudge}. {Although the smudge tool could be exploited for arbitrary images, during painting, users use it most on color regions {to iteratively} create desired shading effects.} With the smudge tool,} users may draw strokes with manually adjusted radii to cover parts of a painting for smudging. However, this traditional tool is not color region-aware or shape boundary-aware during smudging. All areas covered by the strokes are smudged evenly, thus often leading to undesired artifacts. For example, intended sharp edges might be smoothed out, or unwanted colors covered by the strokes might mix into the blending result and thus disturb the creation of expected shading effects {(Fig.~\ref{fig:prepost})}. 

To avoid such artifacts, professionals often manually create masks with the selection tool or {decompose a painting into multiple layers with {the layer creation tool}}, so that the target areas for smudging can be separated from the other areas of the painting. Although {such practice} {is a conventional procedure for professionals to achieve desired blending results, it is challenging and time-consuming for novices. As confirmed by our formative study, frequent switching between layers in a multi-layer mode is laborious for users. Users found that smudging in a single layer made it hard to fulfill natural and complex shading effects without the help of multiple layers. Last, novice users found it hard to select an appropriate size of a smudge brush during color smudging.}

\begin{figure}[t]
\centering
  \includegraphics[width=0.8\linewidth]{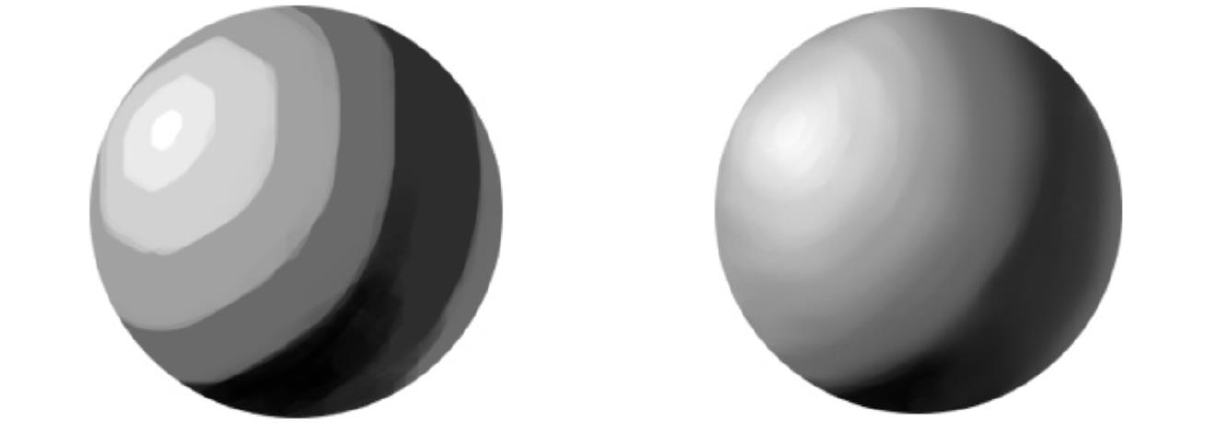}
\caption{
Left: hard shading effects. Right: smooth shading effects.} 
\label{fig:shading}
\end{figure}

To relieve the learning and interaction burden for users and enable them to create complex and natural shading effects, we propose a novel smudge tool, \sysName, which recognizes users' intentions based on smudge paths with respect to the regions being covered and dynamically adjusts the brush radius to preserve color boundaries or soft blend color regions (Fig.~\ref{fig:pipeline}). \sysName~is designed based on the following key observations. First, users tend to smudge colors {either around edges or into color regions. Thus, the smudge paths generated by users tend to resemble either regions or boundaries.} Second, users take advantage of masks generated from multi-layer paintings to preserve boundaries during color smudging, while frequent switching between layers is tedious. How to keep sharp edges and create natural shading effects at the same time without using additional layers and masks is valuable to explore. Our proposed tool enables users to smudge color more accurately to create both hard shading effects and smooth shading effects simultaneously \emph{in a single layer} {(Fig. \ref{fig:shading})}, thus empowering users to make creative work more efficiently and intuitively.

We conducted a user study with 12 users (7 novice users, 5 professional users) in two independent tasks to validate the effectiveness and intuitiveness of our system, which integrated the \sysName~tool and other auxiliary functions for painting, e.g.,  traditional smudge tool, scaling, changing brush size, undoing, etc. In addition, we invited 3 professional users to create showcases by using \sysName~to show its creativity and expressiveness.

In sum, we make the following contributions in this work: We conducted a formative study to investigate challenges and explore interaction technology of color smudging. {Based on the findings in the formative study, we propose} a color smudging algorithm, supporting dynamic size-adaptive brushes and real-time region selection for easily creating natural shading effects in a single layer. Then, we implemented a prototype system based on the proposed algorithm and evaluated its effectiveness, efficiency, and expressiveness via a user study. We will release the code of our tool upon acceptance of the paper.

\section{Related Work}
\subsection{Drawing Assistance Tools} 

Several systems assisted users in drawing a shape more accurately by offering visual guidance \cite{williford2019drawmyphoto, dixon2010icandraw, iarussi2013drawing, Lee:1978:TQA:800025.1198348} and even beautifying strokes \cite{su2014ez, fivser2015shipshape, igarashi2007interactive, limpaecher2013real, xie2014portraitsketch, xing2014autocomplete, matsui2016drawfromdrawings}. For example, ShipShape \cite{fivser2015shipshape} and Interactive Beautification \cite{igarashi2007interactive} recognized geometric relations of strokes first and then beautified strokes based on these relationships. DrawFromDrawings \cite{matsui2016drawfromdrawings} and Real-time Draw Assistance \cite{limpaecher2013real} beautified strokes by stroke interpolation and correction vector fields, respectively. EZ-sketching \cite{su2014ez} took advantage of a single reference image to correct strokes locally, semi-globally, and globally. Those works corrected shapes or beautified strokes, while our paper focuses on how to assist users in painting efficiently and creatively without changing any user inputs. The Drawing Assistant \cite{iarussi2013drawing}, DrawMyPhoto \cite{williford2019drawmyphoto}, and iCanDraw \cite{dixon2010icandraw} provided users with visual tutorials generated from a source image to guide users. However, they still concentrated on drawing shapes rather than the painting process. Instead of assisting humans in drawing, several works focus on enabling robots to create paintings. For example, Adamic et al.~\cite{adamik2022fast} proposed a novel general robotic system for creating realistic pencil drawings based on image evolution, and Karimov et al.~\cite{karimov2023robot} introduced a data-driven mathematical model for artistic paint mixing. Compared with these works, our goal is to design a novel tool for easily adding shading effects by inferring users' intentions from the interaction of smudge strokes and the covered regions.

\subsection{Image and Sketch Colorization}
Early work in colorization transferred color from a source image into a grey-scale image \cite{welsh2002transferring, ironi2005colorization}. Levin et al. \cite{levin2004colorization} colorized images by propagating a few user-specified color scribbles spatially and temporally. With the advance of artificial intelligence, an increasing number of data-driven methods were adopted to colorize images with reference color images \cite{lu2020gray2colornet, he2018deep}, color strokes \cite{xiao2019interactive}, color dots \cite{zhang2017real}, or text descriptions \cite{zou2019language, weng2022code}. In contrast, our paper focuses on sketch colorization {by smudging} instead of image colorization.

Compared with image colorization, it is more challenging to colorize sketches lacking greyscale features. (or line {artwork}) \cite{zhang2018two}. Qu et al. \cite{qu2006manga} and LazyBrush \cite{sykora2009lazybrush} propagated constant colors over regions of sketches with pattern continuity and drawing styles. Recently, machine learning methods were applied {to} sketch colorization by reference sketches \cite{akita2020colorization}, color strokes \cite{zhang2021user}, text descriptions \cite{kim2019tag2pix} or color dots \cite{zhang2018two}. However, those sketch colorization approaches generate flat filling colorization by using color information offered by users in the first step, which limited {the} creativity of artwork and lacked {detailed control for creating complex} shading effects. Our work {concentrates} on assisting users in painting from the beginning efficiently and intuitively, empowering users to make more creative artwork.

\subsection{Shading, Smoothing, and Smudging}
PlayfulPalette \cite{shugrina2017playful} and nonlinear color triads \cite{shugrina2020nonlinear} allowed users to create hard shading effects with nonlinear color palettes. Sochorov\'{a} et al.\cite{sochorova2021practical} integrated the Kubelka-Munk model into the RGB color representation to more realistically depict blended pigments in digital painting. In contrast to these works, our work focuses on how to create smooth shading effects. Bi et al. introduced an image flattening method \cite{bi20151} based on the L1 norm to generate edge-preserving smoothing effects. Similarly, L0 gradient minimization \cite{xu2011image}, bilateral grid \cite{chen2007real}, EAP \cite{zhang2020erasing}, and truncated total variation \cite{dou2017image} were adopted to smooth images. Instead of generating smooth shading effects on an image or a sketch, we focus on how to create smooth shading effects on a painting.

For relatively simple shading effects, shading curves \cite{orzan2008diffusion, lieng2015shading} {might be} used to cast shading effects into line drawings directly at once. For {more complex shading effects, users might create and revise them} step by step by exploiting different types of brushes, such as a smudge brush, a blur brush, or a gradient-domain brush \cite{mccann2008real}. Since the blur tool, in general, softens the edges between color regions to smooth painting results, it is not suitable for creating complex shading effects inside color regions, as shown in Fig. \ref{fig:blurvssmudge}. In contrast, the smudge tool generates natural shading effects not only along edges of color regions but also inside color regions \cite{kwak2013adaptive}. 

Generally, the smudge tool from commercial software has complicated parameter settings to create well-controlled shading effects. The parameter tuning is time-consuming and not friendly for novice users. In addition, smudging {with} the smudge tool without any limit easily {causes} blurred color region boundaries (Fig. \ref{fig:prepost}). Such effects might sometimes be against users' intentions. Therefore, to perform color smudging in desired color regions, users tend to smudge color regions under a mask created with a selection tool in multiple layers\cite{photoshop}. This overwhelming, complex process is challenging, {especially for novice users.} {Jr et al. \cite{olsen2008edge} proposed an edge-respecting brush spreading painting effects according to the edges and texture of images. Lazy Selection \cite{xu2012lazy} presented a quick selection system by matching user inputs with geometry rules, thus enabling users to quickly select elements corresponding to their intention. The Bubble Cursor \cite{grossman2005bubble} dynamically adjusted its selection range based on the proximity of nearby targets, facilitating easier area selection.} Inspired by these works, we propose a smudge tool \sysName~in order to allow users to create desired natural shading effects in a single layer based on geometry rules extracted from the edge information. Our paper introduces a new region selection algorithm and offers a dynamic size-adaptive brush, thus empowering users to create region-aware shading effects in their desired color regions more efficiently and intuitively.

\section{Formative Study}
We conducted a formative study to investigate users' behaviors when they were using a traditional smudge tool. {The study included two tasks. The first task was to reproduce paintings from scratch, and the second task was to reproduce paintings from given flat-filled paintings. The observations and feedback motivated us to devise the proposed \sysName~tool. }

\subsection{User Study}

\textbf{Participants and Apparatus.} 
We recruited 6 participants (P1 to P6) for Task 1 and 12 participants (P7 to P18) for Task 2. In Task 1: P1 to P3 were professionals or amateurs with experiences in digital painting {(P1: {a} professional, P2 and P3: {two} amateurs)}; {P4 to P6} were novices who were not familiar with digital painting. {P1 to P4} were familiar with traditional painting. In Task 2: P7 to P11 were professionals, while P12 to P18 were novices, who were not familiar with digital painting. We provided three stylus input devices to the participants: a Wacom digital tablet, an iPad Pro with an Apple pencil, and a Microsoft Surface Book 3 with a Surface Slim pen. Photoshop, Clip Studio Paint, and Procreate were adopted to create paintings. In both tasks, the participants could freely choose any input device. In {T}ask 1, we offer{ed} Clip Studio Paint, a mainstream commercial painting application with over 30 million users, as our painting software. In Task 2, they could freely select software for the study.

\begin{figure}[t]
\centering
\includegraphics[width=0.98\linewidth]{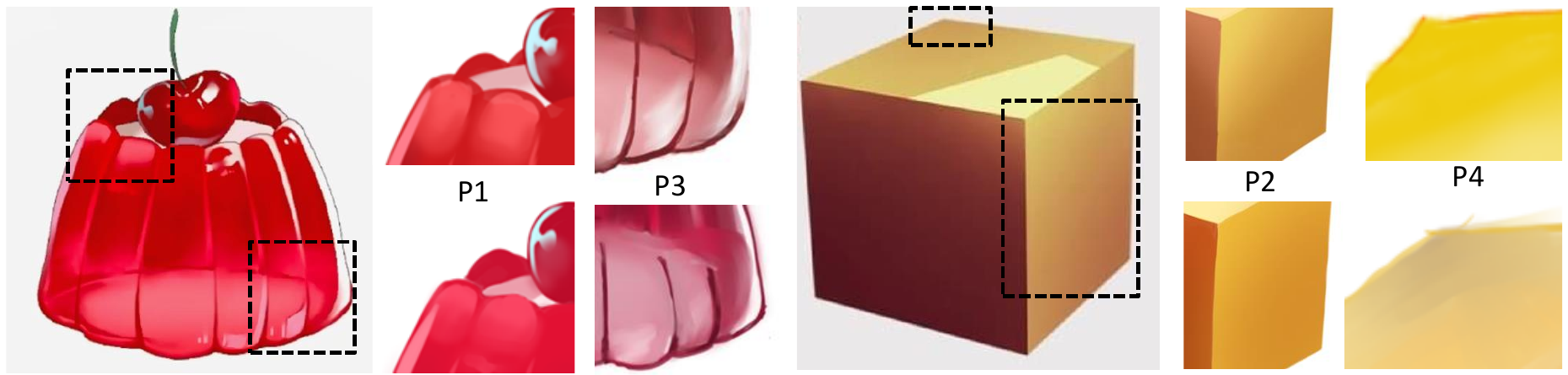}
\caption{{The top and bottom rows show the paintings created by the same participants in the multi-layer mode and the single-layer mode, respectively. The first and fourth (from left to right) columns give the corresponding reference images.}}
\label{fig:allpaintings}
\end{figure}

\textbf{Tasks.} 
The first task was to reproduce two target digital paintings with rich shading effects, i.e., the cube and the jelly cake (Fig.~\ref{fig:allpaintings} (Left)), with {a fingertip tool {({for blurring and dragging} colors along the {moving} directions of the cursor, like the traditional smudge tool)}, and multi-layer functions provided by Clip Studio Paint.} It had two configurations, i.e., multi-layer and single-layer configurations. In the multi-layer configuration, the participants were allowed to use the multi-layer function provided by the software to decompose a painting being edited into different parts and place individual parts on different layers. Professional artists have widely adopted this configuration in digital painting creation. In the single-layer configuration, the participants were required to reproduce the target paintings using a single layer. In both configurations, the participants {were instructed to freely paint to reproduce each target drawing within one hour.} To facilitate the participants in creating desired shading effects, we allowed them to pick up colors from the target paintings directly. Before the task, we offered each participant a warm-up session, including a 60-minute session with a tutorial on {the fingertip tool} and free practice with technical support. All the participants felt confident in using the {fingertip} tool after this session. In addition, we allowed the participants to take breaks after finishing each drawing to avoid fatigue. Excluding the breaks and the tutorial session, the task lasted around 2.5 hours on average for each participant. After the task, we conducted a 30-minute semi-structured interview with the participants to collect their feedback on color smudging and painting.

\begin{figure}[t]
\centering
\includegraphics[width=0.98\linewidth]{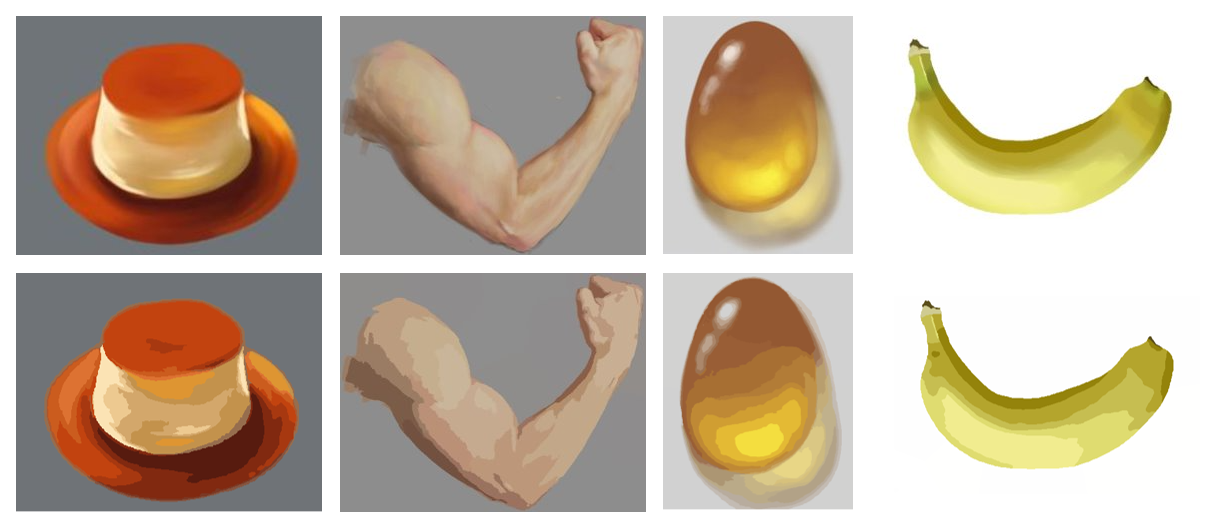}
\caption{{The top and bottom rows show the target paintings and source paintings in Task 2.}}
\label{fig:refimg}
\end{figure}

In Task 2, we offered flat-filled paintings as the source and reference paintings as the target. We required the participants to paint and smudge based on the flat-filled paintings with {the traditional smudge tools, i.e., the smudge tool in Photoshop and Procreate, and the fingertip tool in Clip Studio Paint.} The participants were allowed to choose two of four paintings (shown in Fig.~\ref{fig:refimg}) and reproduce them from the corresponding reference paintings. In addition, they were allowed to change the brush size via shortcuts from a keyboard and the pressure sensitivity of the brush. Like the first task, we offered each participant a warm-up session and allowed them to take breaks. It took about 1 hour for each participant to finish this task. After the task, the study participants answered a questionnaire and attended a 15-minute semi-structured interview. In both tasks, the order of the target drawings and the configurations were counterbalanced by a Latin square design to minimize possible learning effects. 

\subsection{Observations} 
All the participants completed the tasks. We observed the processes of reproducing the target paintings, collected their feedback after the task, and obtained the following interesting findings about using the smudge tool. In Task 1, on average, they used 6.17 layers (std = 2.48) for the cube and 7.67 layers (std = 4.64) for the jelly cake in the multi-layer configuration. The multi-layer configuration was preferred by P1, i.e.{,} the professional participant.  In this configuration, she could easily and efficiently create desired and complex shading effects, such as reserving sharp boundaries and creating smooth shading effects, {as shown in Fig. \ref{fig:allpaintings}}. In contrast, the novice participants preferred the single-layer configuration since {painting with layers required professional skills.}  According to a paired t-test ({$p < .05$}) conducted on the time for completing each task in the single-layer configuration (short for SC) and the multi-layer configuration (short for MC), we observed that the participants spent significantly less time in SC (SC: 35.5 mins vs MC: 39.75 mins on average, \emph{p = .0352}). Although the professional participant preferred a multi-layer configuration, she agreed that painting in a single-layer configuration was more convenient.

In Task 2, the participants used both long and short smudge strokes. When they drew a long stroke, their interaction purposes might be updated or switched dynamically. For instance, P12 commented that although he used a long smudge stroke, he only focused on the current partial smudge strokes. The direction of smudge strokes varied a lot, depending on the users' intentions. They might draw multiple short strokes parallelly or draw a long stroke back and forth. P8 said that smudging back and forth generated zigzag shading effects, and it was harder for him to control, so he preferred to exploit parallel smudge strokes. In contrast, P11 thought that dense short strokes in a zigzag direction created more natural shading effects. In addition, during color smudging, the participants smudged both inside regions and around edges based on their intention as shown in Fig. \ref{fig:smudgestyle}. Smudging around edges offers evener shading effects between two colors from the two sides of the edges while smudging into regions tends to create the effect that one color invades another color. P8 commented that she might smudge into regions when creating more abrupt shading effects and smudge around edges to generate smooth color gradients. P14 commented, ``It is better to implicitly change the brush size instead of explicitly using shortcuts from the keyboard.'' On the other hand, for implicit brush size selection via pressure sensitivity, some users, especially the novices, felt it challenging to change the brush size accurately via pressure sensitivity because it required precise pressure control. Last, we found that the users tended to use a larger brush size first on a target region to reduce the drawing efforts and exploited a small size later to refine shading effects.

\subsection{Challenges}
We summarized several challenges for achieving natural shading effects by grouping and analyzing the participants' behaviors and answers.

\textbf{Improper Brush Size.}
Many participants, especially novices, mentioned that choosing the appropriate brush size was challenging. During color smudging, the participants frequently created undesired artifacts because of choosing an improper brush size and then had to undo or redo the operation. For example, some necessary sharp edges might be smoothed out, such as the bottom of the jelly cake (Fig.~\ref{fig:allpaintings}), or unwanted colors might be covered by smudge strokes, resulting in undesired shading effects. What's more, they complained about frequent switches between different brush sizes. Therefore, how to generate an appropriate dynamic brush size automatically and implicitly without frequent adjustment of the brush size is valuable to explore.

\textbf{Frequent Switches and Selection between Layers.}
During color smudging, users tended to smudge different parts in the corresponding layers or select smudging regions with masks to get different shading results in different regions. For instance, the participants (P2 and P3) created two different layers for smudging the cherry and the white surface of the jelly cake to keep the edges between those two parts. To create natural shading effects, the participants had to switch layers, exploit the selection function, and use masks frequently. It is vital to explore how to create natural shading effects without additional layer creation and selection.

\begin{figure}[t]
\centering 
\includegraphics[width=0.95\linewidth]{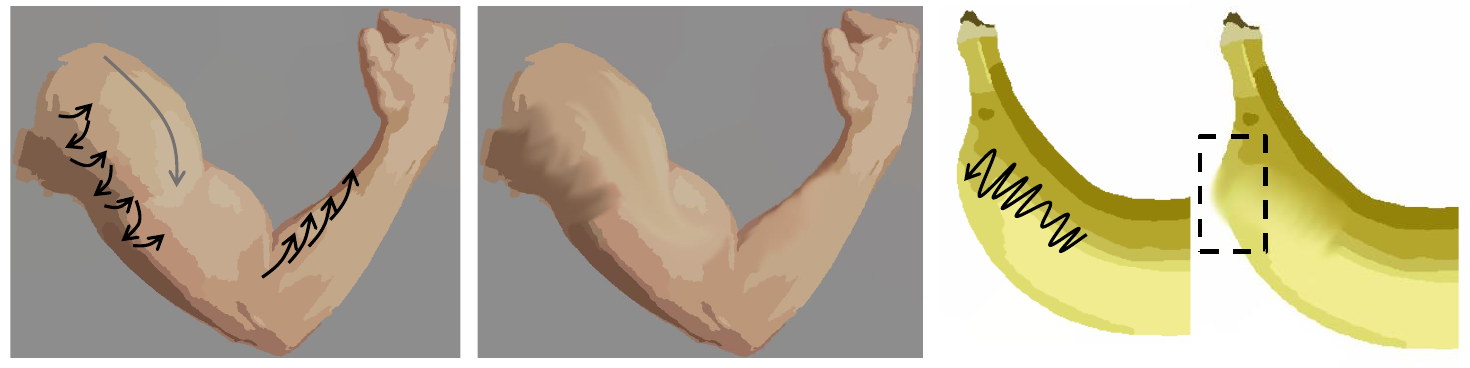}
\caption{From left to right, columns 1 and 2 illustrate smudge strokes and the corresponding shading effects. Here, a gray arrow indicates a smudge stroke into regions, while two sequences of black arrows show zigzag and parallel smudge strokes around edges. Columns 3 and 4 indicate over-smooth artifacts.}
\label{fig:smudgestyle}
\end{figure}

\textbf{Over-Smoothing.}
Over half of the participants found it hard to create natural shading effects and preserve sharp boundaries. For example, P16 commented that he could recognize the approximate color gradients and know where to smudge, but it was still challenging for him to create smooth shading effects, not to mention keeping sharp boundaries at the same time. In most cases, it is common for users to smooth out too much. As a result, all the sharp boundaries disappeared, including some that needed to be preserved, such as the boundary between the banana's top and side (Fig. \ref{fig:smudgestyle}).

\section{{Methodology}}\label{sec:alg}

\subsection{Solutions} 
To tackle the aforementioned challenges in the formative study, we put forward several solutions based on users' behaviors and intentions. Then we devise \sysName, a novel smudge tool that automatically predicts users' intentions according to their drawn smudge strokes based on the solutions. With this tool, users can efficiently create desired blending effects in the single-layer configuration without performing delicate operations.

\textbf{Solution to Improper Brush Size.}
Improper brush sizes impede users from creating natural shading effects efficiently. Our tool generates a size-adaptive dynamic brush based on current smudge strokes, which empowers users to create shading effects more efficiently{, thus reducing} the drawing effort.

\textbf{Solution to Frequent Switches and Selections between Layers.}
Smudge strokes reveal users' intentions. As mentioned before, users smudge around edges or into regions. Selecting the target smudging regions with simple criteria, e.g., whether a smudge stroke covers a region or its centerline crosses a region, may lead to undesired results, since these criteria are sensitive to users' precise operations. Thus, to select the target smudging regions according to users' intentions and provide users easy control, our algorithm compares the similarity between the smudging color regions with smudge strokes in edges and regions, defined as region resemblance and edge resemblance. In this selected target smudging color region, users can simultaneously smudge color and preserve edges without additional layer creation and selection operations. We do not consider the color information for region selection, since users may blend regions in similar or distinct colors, depending on specific painting tasks. Exploiting such information may introduce additional variations when users perform smudge operations in different painting tasks.

\textbf{Solution to Over-Smoothing.}
It is common for users to {create over-smoothing artifacts} during color smudging {with the traditional tools} (Fig. \ref{fig:allpaintings} and Fig. \ref{fig:smudgestyle}). To solve this issue, by relying on a dynamic mask generated on selected smudging regions and a real-time region-splitting algorithm, our tool enables users to preserve the edges of color regions, leading to a balance where smoothing effects are achieved without sacrificing the sharpness of edges. This ensures that there is no over-smoothing, maintaining sharp edges of color regions.

\subsection{{Algorithm}}
\textbf{Problem Formulation.} According to {the formative study}, predicting users' smudging intentions is equivalent to recognizing the target regions to be smudged from a stroke. We observe that colors differentiate target regions and unwanted regions. Therefore, instead of directly recognizing the target regions, we adopt a two-step strategy, i.e., extracting a set of color regions and then selecting these regions. These color regions can be obtained directly for a painting created by flat filling. For a general painting without well-defined color regions, we split the painting and extract color regions using the MeanShift method {based on RGBXY information}~\cite{comaniciu2002mean}. We then select the target regions from them according to the drawn stroke. According to {the observations from the formative study}, users' intentions may update during the stroke drawing, therefore the target color regions are dynamically selected with partial strokes. 

\begin{figure}[t]
\centering
  \includegraphics[width=0.9\linewidth]{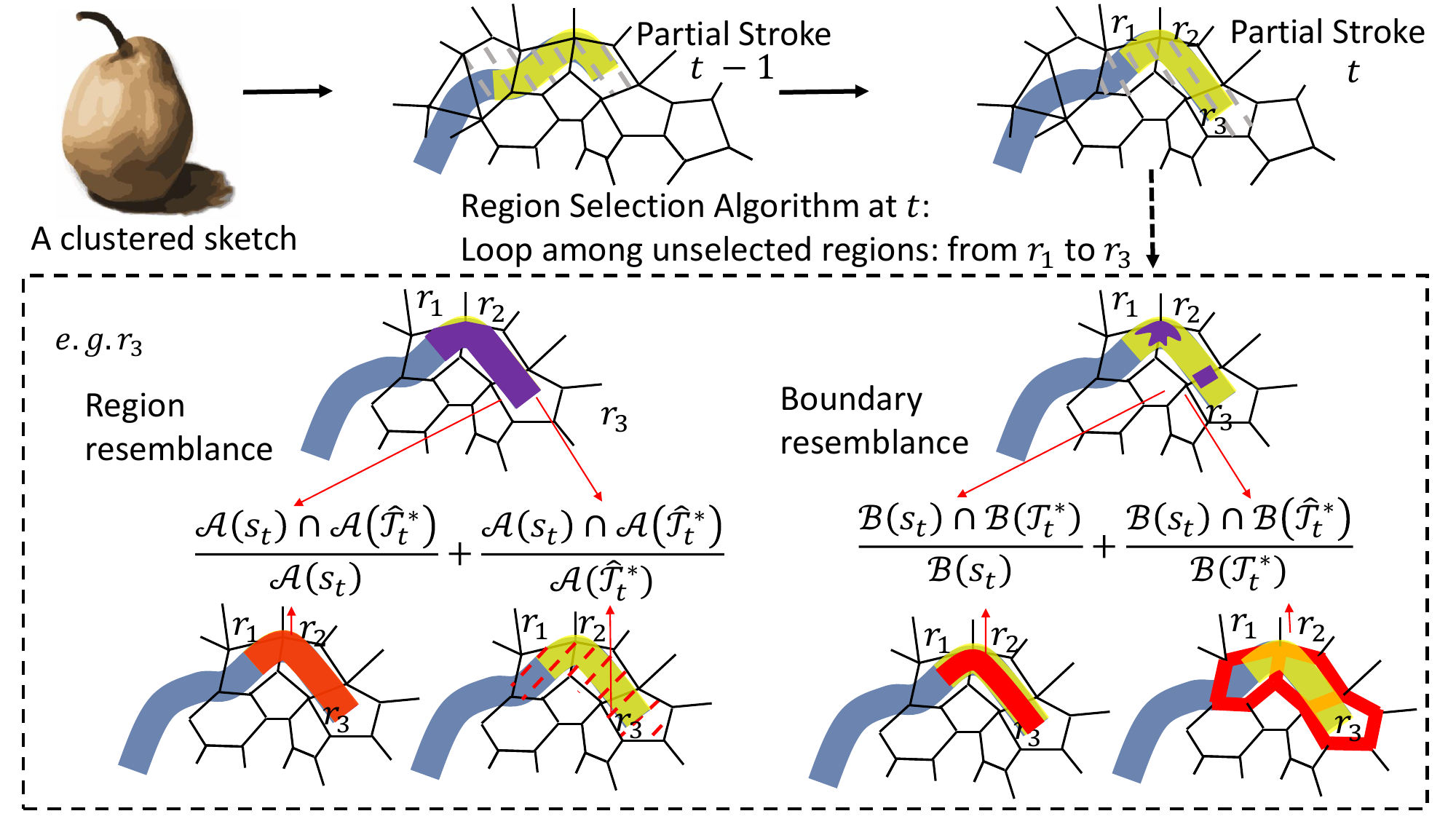}
\caption{The pipeline of the proposed system using region resemblance and boundary resemblance. Aiming at the flat-filled color regions or clustered color regions, we extract a partial stroke (represented in yellow) from a smudge stroke (represented in blue) and then conduct a region selection algorithm on the current color regions, covered by the partial stroke. The red dashed lines or scribbles show the denominators of Equation~(\ref{eqn:reg_res}) and (\ref{eqn:bdy_res}), and the purple regions show the numerators.}

\label{fig:pipeline}
\end{figure}

\textbf{Dynamic Region Selection with Partial Strokes.}
Given a painting composed of a set of color regions, we denote it as $\mathcal{R}$ and $r\in\mathcal{R}$ is one of its color regions. When users draw a stroke, at each timestamp $t$, we consider the partial stroke $s_t$ that is most recently drawn and with a fixed length $l$ and width $w$ for the region selection (see Fig.~\ref{fig:pipeline}). This length $l$ and width $w$ are device-dependent; we will discuss them later. At the beginning of the stroke {drawing, i.e., $t=0$, $s_0$ contains} only one stroke point. During the drawing, the partial stroke $s_t$ covers a subset of $\mathcal{R}$, which we term the covered regions and denote as $\mathcal{C}_t\subseteq\mathcal{R}$. However, some of the covered regions are not the target ones. {The observations concluded in the formative study} suggest that the partial stroke should resemble the target regions or their boundaries, as illustrated in Fig. \ref{fig:pipeline}. A straightforward method would be enumerating the combinations of the covered regions and finding the one that best resembles the partial stroke $s_t$. However, this will lead to a combinatorial explosion, which might prevent a real-time smudge experience. In addition, as the timestamp changes, the target regions estimated by different partial strokes, e.g., $s_{t-1}$ and $s_t$, may change abruptly. This will result in unstable target regions.

\begin{figure}[t]
\centering
  \includegraphics[width=0.98\linewidth]{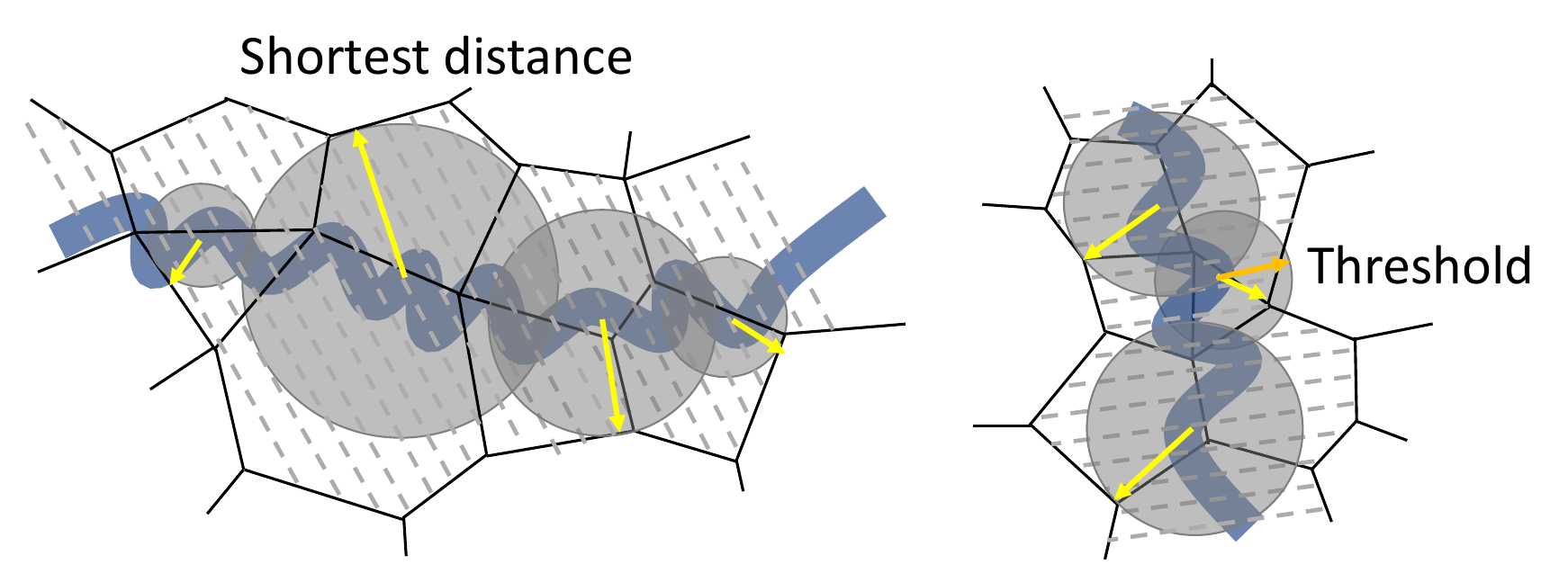}
\caption{{Gray circles show how a brush size changes during color smudging. Yellow and orange arrows indicate the shortest distance and the threshold, respectively.}}
\label{fig:dynamicbrush}
\end{figure}

Instead of independently estimating the target regions with the partial stroke $s_t$ at each timestamp $t$, we devise an algorithm to update the estimated target regions by adding and removing color regions. At timestamp $t$, we suppose the target region set $\mathcal{T}_{t-1}\subseteq \mathcal{C}_{t-1}$ at the previous timestamp $t-1$ is already estimated. This assumption is reasonable since $\mathcal{T}_0$ is an empty set at the initial timestamp, and the target region sets at the following timestamps can be estimated progressively. With this assumption, we need to estimate the new target region set at timestamp $t$. Since the partial stroke changes from $s_{t-1}$ to $s_{t}$, the covered region set also changes. Recall that the covered region set is composed of the color regions covered by the partial stroke, as explained in the previous paragraph. We first update the target region set by removing the uncovered regions, i.e., $\hat{\mathcal{T}}_{t} = \mathcal{T}_{t-1}\cap \mathcal{C}_{t}$. $\hat{\mathcal{T}}_{t}$ is the base target region set, which is a subset of the final target region set $\mathcal{T}_{t}$. That is, once a color region is selected, it remains a target region until it is not covered by the partial stroke.  This guarantees the stability of the target regions. After obtaining $\hat{\mathcal{T}}_{t}$, we further update it by adding the covered but not selected regions. To guarantee a smooth transition between target regions at adjacent timestamps, we allow at most one region to be added at each timestamp. That is, $\mathcal{T}_{t}$ could be $\hat{\mathcal{T}}_{t}$ or $\hat{\mathcal{T}}_{t}^i=\hat{\mathcal{T}}_{t}\bigcup\{r_i\}$, where $r_i\in\mathcal{C}_{t}\setminus\hat{\mathcal{T}}_{t}$. We consider $\hat{\mathcal{T}}_{t}$ and $\{\hat{\mathcal{T}}_{t}^i=\hat{\mathcal{T}}_{t}\bigcup\{r_i\}:r_i\in\mathcal{C}_{t}\setminus\hat{\mathcal{T}}_{t}\}$ 
as the candidate target region sets.

\begin{figure*}[ht]
\centering
  \includegraphics[width=0.95\linewidth]{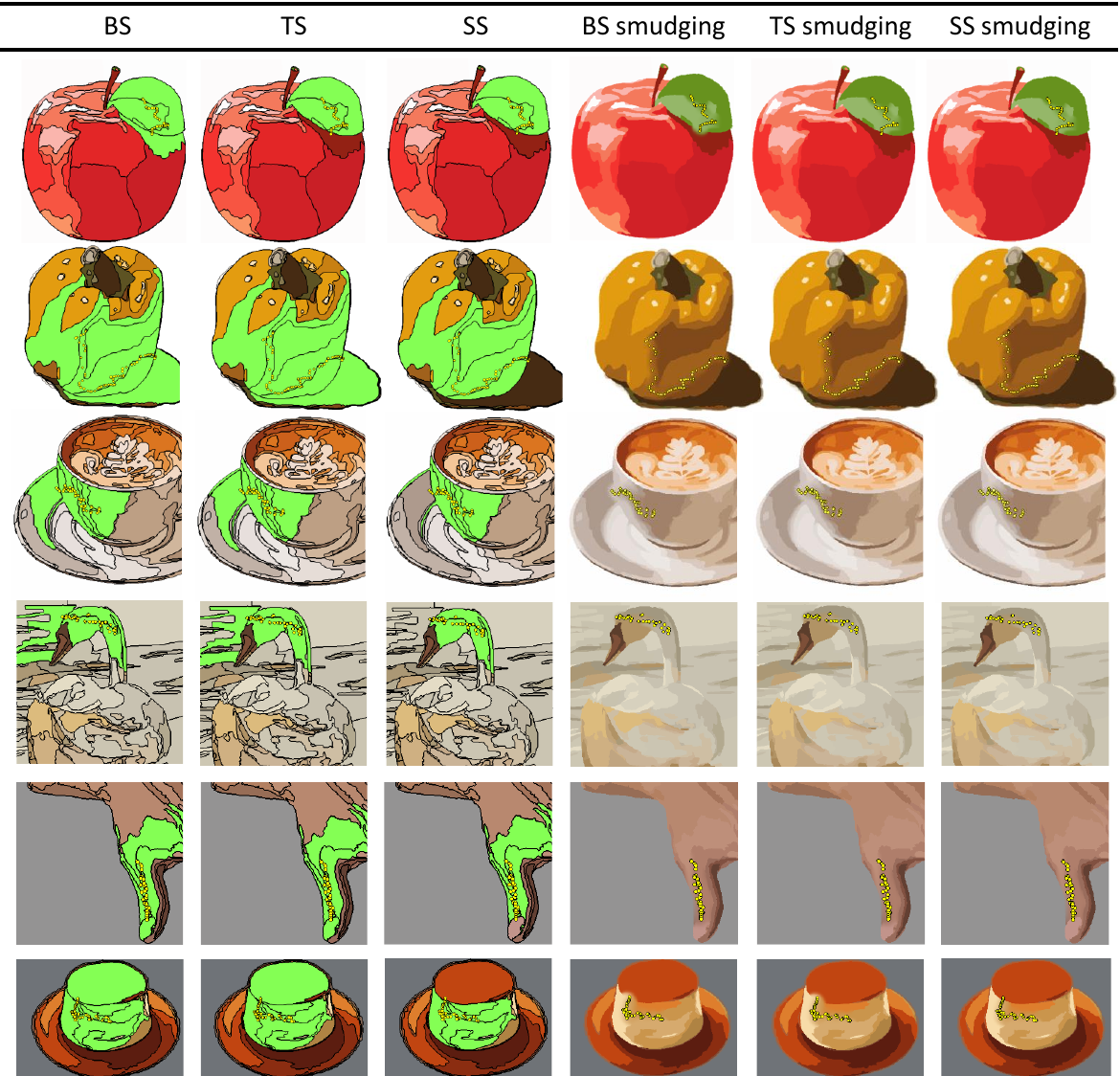}
\vspace{-10pt}
\caption{{Columns 1, 2, and 3 (from left to right) show results of region selection by using the BS, TS, and SS, respectively, and columns 4, 5, and 6 indicate the subsequent smudging results based on each region selection algorithm. Yellow dots or red dots in each figure indicate the smudge path.}
}
\label{fig:regionselectionres}
\end{figure*}

Among all these candidate target region sets, we need to find the one that best resembles the partial stroke $s_{t}$. {According to the observations from {Task 2 in the formative study}, users smudge either around edges to create smooth shading effects or into regions to create more invaded shading effects.} These two intentions are revealed by the region resemblance and the boundary resemblance respectively. Since users' intentions cannot be determined beforehand, we combine the region resemblance and boundary resemblance to define a resemblance score between the partial stroke $s_t$ and a candidate target region set (Fig. \ref{fig:pipeline}). 

The region resemblance is measured with two terms, i.e., the \emph{region coverage} and the \emph{stroke inclusion in the region}. Given a candidate target region set $\hat{\mathcal{T}}_{t}^i$, the region coverage is defined as the percentage of this candidate that is covered by the partial stroke. A high region coverage indicates a high possibility of this candidate being the target. The stroke inclusion in the region is defined as the percentage of the partial stroke that is included in this candidate. A high stroke inclusion implies that users' effort is respected and thus this candidate is likely to be their intention. Based on this discussion, we define the region resemblance score $R_r$ as:
\begin{equation}\label{eqn:reg_res}
    R_r(s_{t}, \hat{\mathcal{T}}_{t}^*) = \frac{\mathcal{A}(s_{t})\cap\mathcal{A}(\hat{\mathcal{T}}_{t}^*)}{\mathcal{A}(\hat{\mathcal{T}}_{t}^*)} + 
    \frac{\mathcal{A}(s_{t})\cap\mathcal{A}(\hat{\mathcal{T}}_{t}^*)}{\mathcal{A}(s_{t})},
    \label{eq1}
\end{equation}
where $\mathcal{A}(\cdot)$ means the area point set, {i.e., the set of pixel points covered by the partial stroke or contained by} the candidate target region set; $\hat{\mathcal{T}}_{t}^*$ is a candidate target region set; the index $*$ is empty or $i$, corresponding to $\hat{\mathcal{T}}_{t}$ or $\hat{\mathcal{T}}_{t}^i$. 

The boundary resemblance is measured by another two terms, i.e., the \emph{boundary coverage} and the \emph{stroke inclusion on boundary}. These two terms are defined similarly to the region coverage and the stroke inclusion in region. Therefore, we define the boundary resemblance score $R_b$ as:
\begin{equation}\label{eqn:bdy_res}
    R_b(s_{t}, \hat{\mathcal{T}}_{t}^*) = \frac{\mathcal{B}(s_{t})\cap\mathcal{B}(\hat{\mathcal{T}}_{t}^*)}{\mathcal{B}(\hat{\mathcal{T}}_{t}^*)} + 
    \frac{\mathcal{B}(s_{t})\cap\mathcal{B}(\hat{\mathcal{T}}_{t}^*)}{\mathcal{B}(s_{t})},
    \label{eq2}
\end{equation}
where the definition of $\mathcal{B}(\cdot)$ depends on the parameter in the brackets. For instance, $\mathcal{B}(s_{t})$ represents the expansion of boundary point set, {i.e., the expanded (10 pixels) set of pixel points from the boundary of the candidate target region set,} and $\mathcal{B}(\hat{\mathcal{T}}_{t}^*)$ indicates the expansion of bone point set, {i.e., the expanded (5 pixels) set of pixel points from the bone of the partial stroke.} Note that $\mathcal{B}(\hat{\mathcal{T}}_{t}^*)$ is the union of the extended boundary points of the regions in $\hat{\mathcal{T}}_{t}^*$ (see Fig.~\ref{fig:pipeline}).

Then the resemblance score between a candidate target region set $\hat{\mathcal{T}}_{t}^*$ and the partial stroke $s_{t}$ is defined as:
\begin{equation}\label{eqn:resemblance}
    R(s_{t}, \hat{\mathcal{T}}_{t}^*) = \alpha R_r(s_{t}, \hat{\mathcal{T}}_{t}^*) + \beta R_b(s_{t}, \hat{\mathcal{T}}_{t}^*),
\end{equation}
where $\alpha$ and $\beta$ are weights controlling the effects of the region and boundary resemblance. We set {$\alpha = 0.3$} and {$\beta = 0.7$} in our experiments. Then the final target region set $\mathcal{T}_{t}$ at timestamp $t$ is the candidate with the largest resemblance score. Our algorithm is effective in avoiding selecting unwanted regions. On the other hand, it requires users to pay more effort when they want to include a new region in the target, since the base target region set $\hat{\mathcal{T}}_{t}$ tends to remain the best candidate. To achieve a balance between deselecting undesired regions and selecting desired regions, we treat the base target region set in a different way. Specifically, we first find the best candidate $\hat{\mathcal{T}}_{t}^k$ from $\{\hat{\mathcal{T}}_{t}^i=\hat{\mathcal{T}}_{t}\bigcup\{r_i\}:r_i\in\mathcal{C}_{t}\setminus\hat{\mathcal{T}}_{t}\}$, and then compare its resemblance score with $\gamma R(s_{t}, \hat{\mathcal{T}}_{t})$, where {$\gamma=0.7$} is a balance weight. That is, the final target region set is $\hat{\mathcal{T}}_{t}^k$ if $R(s_{t}, \hat{\mathcal{T}}_{t}^k) > \gamma R(s_{t}, \hat{\mathcal{T}}_{t})$, and $\hat{\mathcal{T}}_{t}$ otherwise.
 
\textbf{Smudging with a Dynamic Mask.}
After estimating the target region set at each timestamp, we allow the smudging occurs only in these selected regions. They thus serve as a dynamic smudging mask. When smudging the colors in these regions, we adopt an existing algorithm~\cite{photoshop} to achieve the blending effects.

\textbf{Correction of Delayed Region Selection.}
The partial stroke $s_t$ with a fixed length $l$ and width $w$ extracted from the smudge stroke decides real-time region selection. However, to some degree, this design causes selection latency. In addition, when confronted with a new target color region, the algorithm succeeds in selecting the region and smudging colors only after the partial stroke enters a part of the region. As a result, it generates a time latency as well. To alleviate the issue, we perform a linear interpolation between the last color smudging position and the current one.

\textbf{Dynamic Brush Width.}
To offer natural shading effects among flat-filled color regions with different sizes, the proposed system provides a dynamic size-adaptive brush during color smudging. After getting the selected color regions, the size of the brush $\lambda$ is automatically decided by the shortest Euclidean distance $\sigma$ (Fig. \ref{fig:dynamicbrush}) between the current position of the smudge stroke and boundaries of color regions. $\sigma$ varies a lot according to different shapes of the selected color regions. For example (Fig. \ref{fig:dynamicbrush}), $\sigma$ might be extremely little when facing a narrow and long region, causing almost no shading effects here. To avoid this circumstance, we set the brush size $\lambda$ with a maximum value between a fixed experience value $\theta$ and $\sigma$: $\lambda = \emph{Max}(\theta, \sigma)$.

\textbf{Partial Stroke Length $l$ and Width $w$.}
Our algorithm starts from uniform point sampling of a sequence of 2D input points returned from input devices as a smudge path. To alleviate a potentially inaccurate input issue from hardware or user input errors, we expand the sampled smudge stroke with a width $w$. In our device (141dpi), we exploit 110 pixels as $w$, which is equal to 2cm adopted by most of users in reality. As for the stroke length $l$, we use 110 pixels equal to 2cm in reality as the length to capture regions.

\section{Evaluation}
To assess the usability and effectiveness of the proposed prototype system, we compared how different brush sizes and region selection algorithms affect shading effect generation and conducted a user study to compare the painting performance between the basic traditional smudge tool and our \sysName~in terms of painting time, operation times, and SUS scores \cite{bangor2009determining}. 

\subsection{Results and Discussion}

\begin{figure*}[ht]
\centering
  \includegraphics[width=0.95\linewidth]{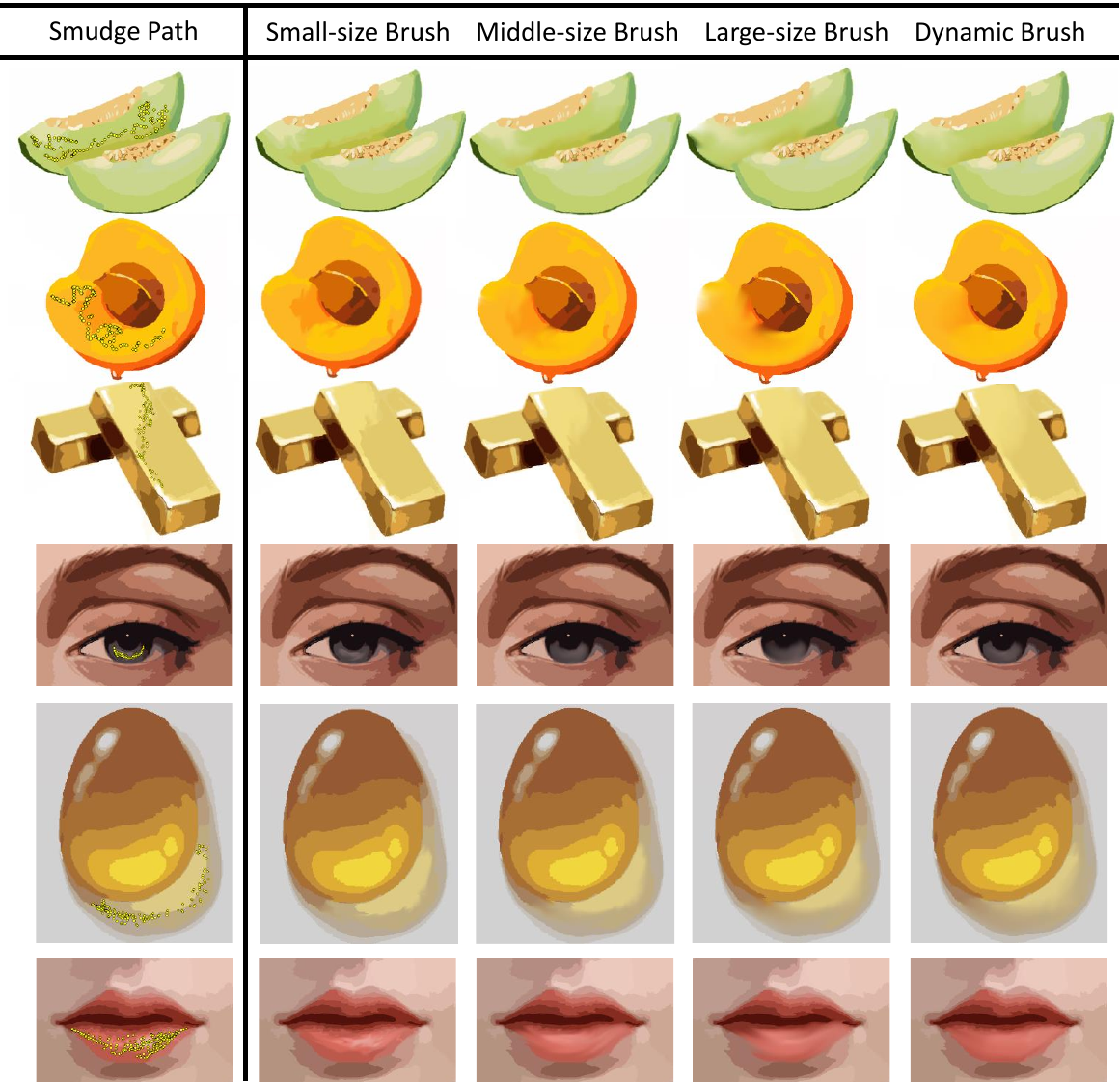}
\vspace{-10pt}
\caption{{Comparisons of results by blending color regions with only a single smudge path but different smudge brush sizes. Column 1 (from left to right) shows the {paintings with flat-filled regions} and the smudge strokes. Columns 2-4 show results by applying a fixed small-size brush (radius: 10 pixels), middle-size  (radius: 25 pixels), and large-size  (radius: 40 pixels) for color smudging, respectively. Column 5 indicates the results by using our dynamic brush without setting up the brush size additionally. All of them are under the same region-based adaptive mask.}}
\label{fig:brushdemo}
\end{figure*}

\begin{table}[t]
\centering
\caption{Performance of the proposed tool on CPU.} 
\setlength\extrarowheight{2.8pt}
\begin{tabular}{@{\extracolsep{4pt}}lcccc}
\hline
\multirow{2}{6em}{Size} & \multicolumn{2}{c}{Region Selection} & \multicolumn{2}{c}{Smudging} \\
\cline{2-3}
\cline{4-5}
& FPS & Milliseconds & FPS & Milliseconds \\
\hline
512 $\times$ 512  & 720 & 1.39 & 48 & 20.83 \\ 
1024 $\times$ 1024 & 240 & 4.17 & 15 & 66.67\\
\hline
\end{tabular}
\label{tab:performance}
\end{table}

\begin{figure*}[!t]
\centering
\includegraphics[width=0.98\linewidth]{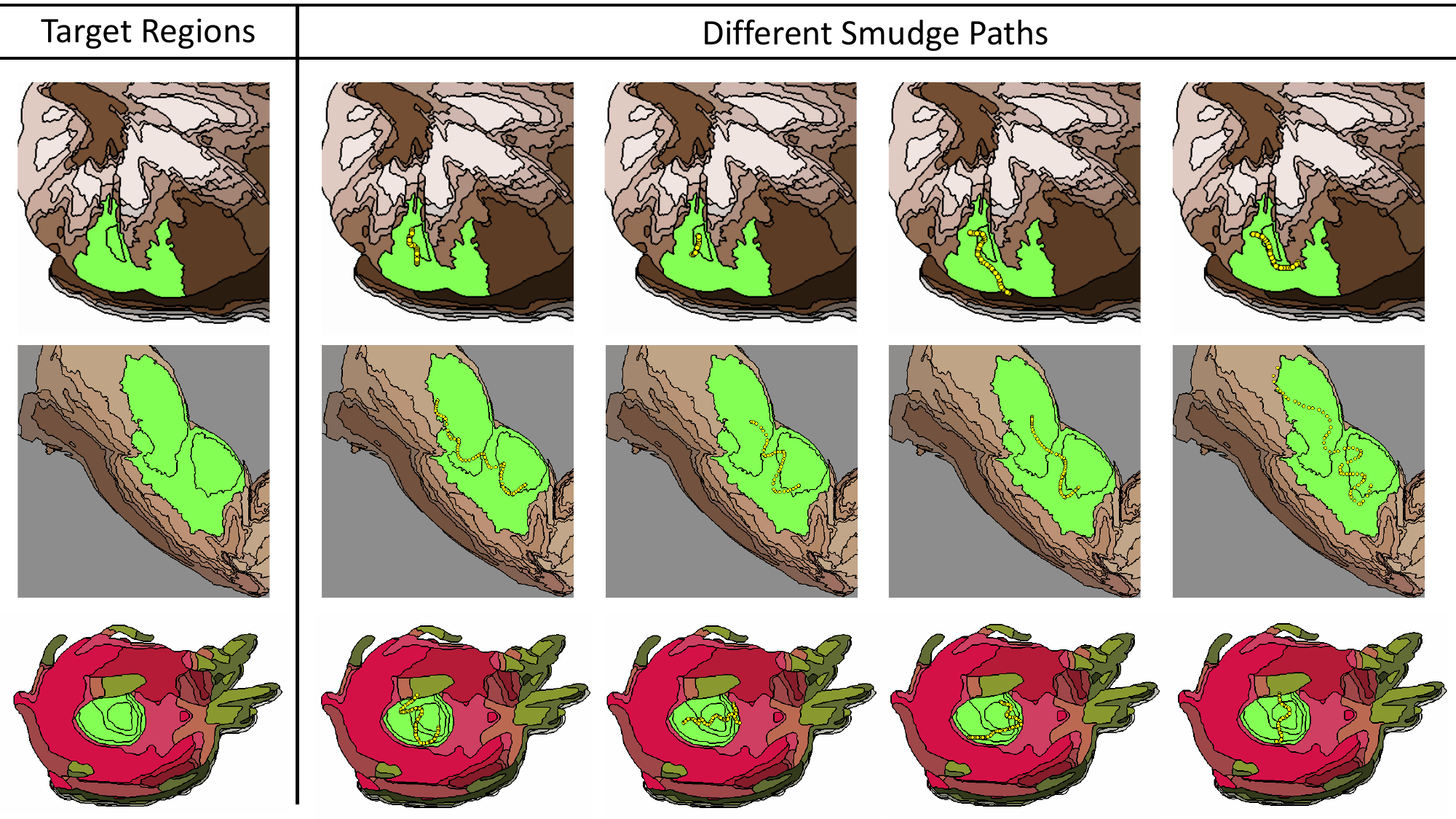}
    \vspace{-10pt}
\caption{{Given target regions  (Column 1), our algorithm works robustly with different smudge paths (Columns 2-5). 
} 
}
\label{fig:robustness}  
\end{figure*}

\begin{figure}[ht]
\centering
  \includegraphics[width=0.9\linewidth]{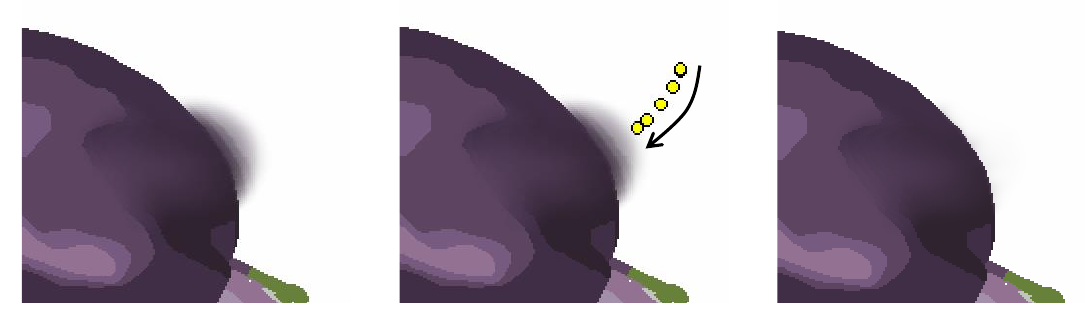}
\caption{After users created artifacts beyond their intention, they could recover edges by smudging color regions again with our region-aware \sysName~tool. Columns 1 to 3 (from left to right) illustrate the artifact, the color smudge operation, and the recovered edge, respectively.
}
\label{fig:edgerecovery}
\end{figure}

\textbf{Regions Selection Results.}
To evaluate the performance of region selection with our proposed algorithm, we compared region selection results generated by: 1). the basic region selection algorithm in commercial software's smudge tool \cite{photoshop} (BS for short); 2). getting {most similar candidates resembling the smudge path from all covered coloring regions $C_{i}$ (we get top 85\% candidates for region selection in our experiments, TS for short)}; 3). the proposed \sysName~algorithm, (SS for short).
As Fig. \ref{fig:regionselectionres} indicates, {the TS algorithm} failed to select continuous regions, thus giving rise to discrete smudging results. As for the edge preservation, {the BS and TS algorithms} could not select regions according to users' intentions stably. In contrast, our algorithm offered accurate continuous region selection results and helped users not to blend colors out of edges. In addition, our region selection results enabled users to recover edges from blending regions, as shown in Fig. \ref{fig:edgerecovery}. {Last, to evaluate the robustness of the region selection algorithm, we tested the algorithm on 40 paintings. As illustrated in Fig. \ref{fig:robustness}, our algorithm selected the target flat-filled regions accurately with different smudge paths.} 

\begin{figure}[t]
\centering
\includegraphics[width=0.95\linewidth]{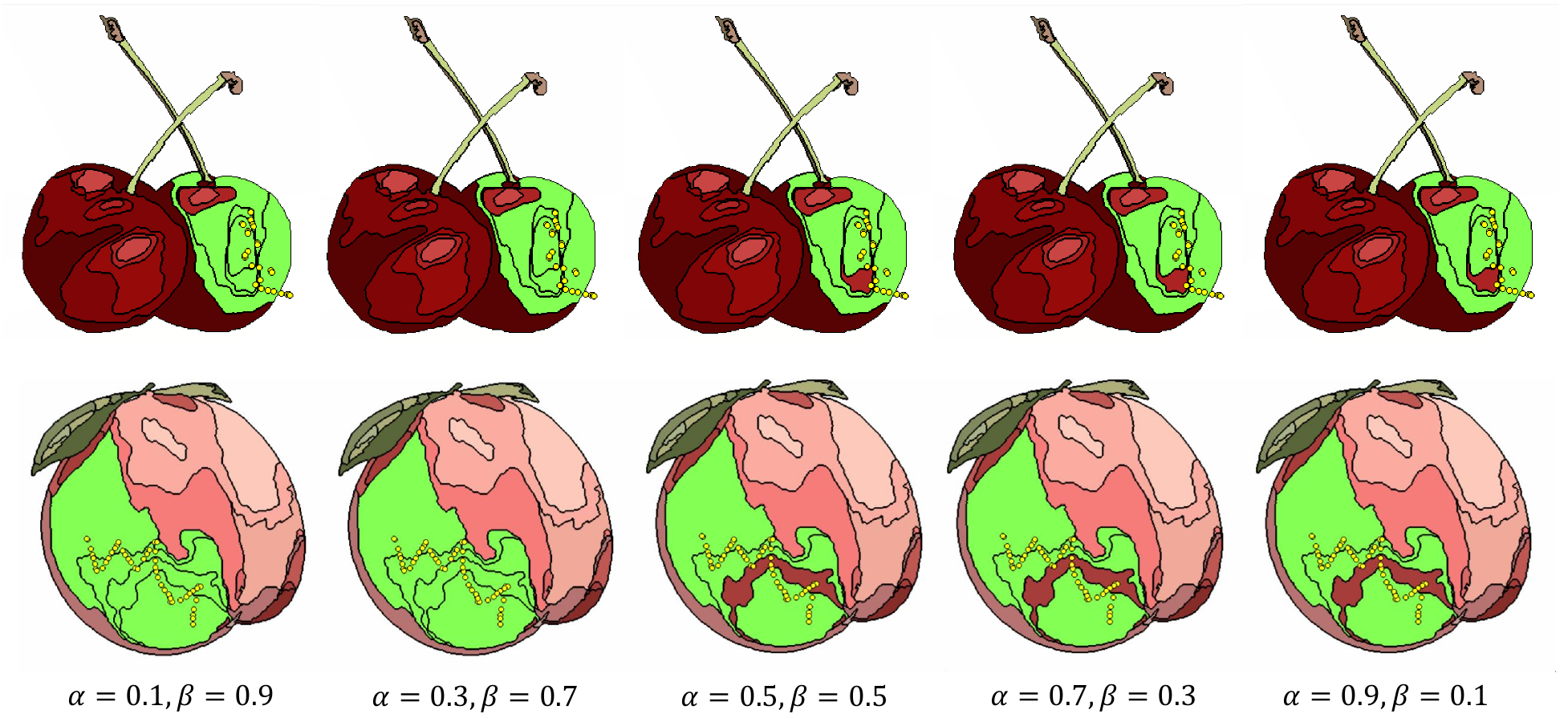}
\caption{Our experiments showed the results of region selection under different parameter values of region and boundary resemblance.}
\label{fig:parameter}
\end{figure}

\begin{figure}[t]
\centering
\includegraphics[width=0.95\linewidth]{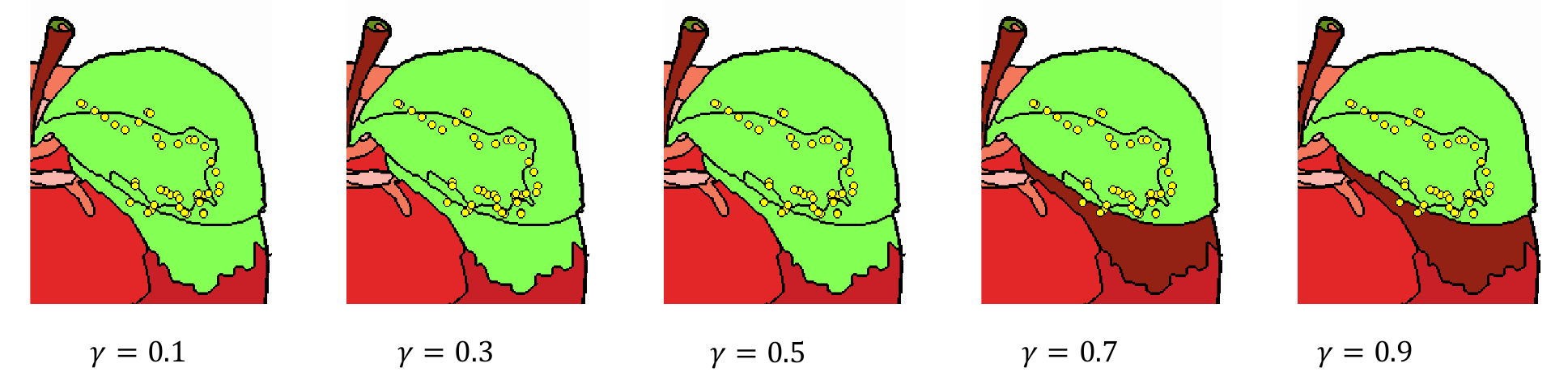}
\caption{
 Adopting different $\gamma$ values influences the results of region selection.}
\label{fig:gammasetting}
\end{figure}

\textbf{Dynamic Brush Results.} To assess the effectiveness of the size-adaptive dynamic brush, under the same adaptive mask generated by region selection, {we compared smudging results by smudge brushes of fixed sizes with those by our dynamic size-adaptive brush. As shown in} Fig. \ref{fig:brushdemo}, the smudge results created by the fixed small brush {lack} natural shading effects in a large range. The fixed large brush {fails} to preserve fine details and {tends} to blend colors outside blending regions against our expectations. While the fixed middle size cannot preserve fine details or create large-scale natural shading effects. As for the proposed dynamic brush, as shown in Fig. \ref{fig:brushdemo}, it naturally creates not only highlight and shadow parts but also fine color gradient changes at all scales.

\begin{figure*}[!t]
\centering
\includegraphics[width=0.98\linewidth]{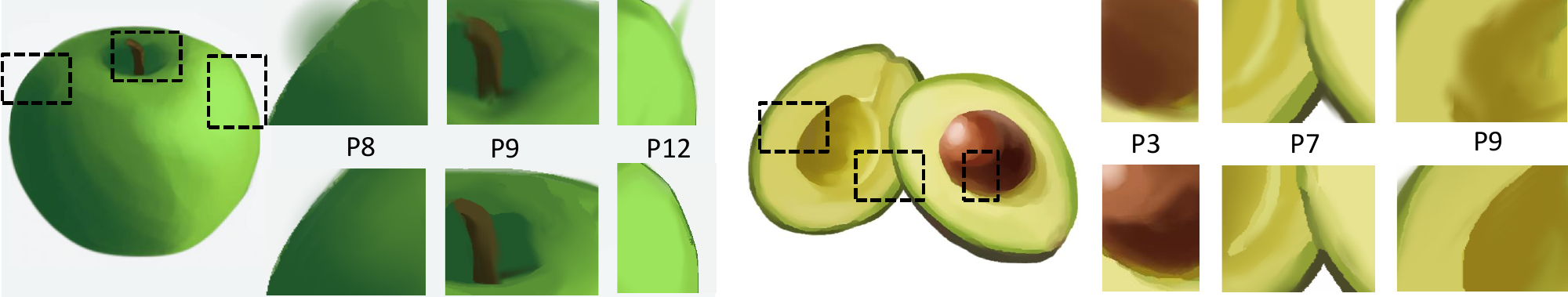}
    \vspace{-10pt}
\caption{Representative results in Study {1}: the results in the top and bottom rows were created by using the basic traditional smudge tool and \sysName~tool, respectively. {The 1st and 5th columns (from left to right) give} the corresponding reference {paintings}. 
}
\label{fig:study1}  
\end{figure*}
\textbf{Parameter Setting {and Performance.}} We {adopt} $\alpha,\beta$ {${\in (0, 1)}$} as resemblance weights and $\gamma {\in (0, 1)}$ as a balance weight in the proposed region selection algorithm (Sec \ref{sec:alg}). As Fig. \ref{fig:parameter} indicates, we {find} that the greater the value of {$\gamma$} is, the {fewer} possible regions are selected, especially after a long smudging path. For instance, if $\gamma$ is equal to $1.0$, users have to make more efforts to select a new region, since this region needs to increase the resemblance score. This becomes difficult when users draw a stroke crossing large regions.
However, the results of setting a low value to $\gamma$, such as $0.5$ in Fig. \ref{fig:gammasetting}, {are} against users' intention{s} as well because the algorithm {selects} almost all covered regions and {does} not keep sharp edges anymore. Thus, we set $\gamma = 0.7$ in our algorithms. As for $\alpha$ and $\beta$, Fig. \ref{fig:parameter} shows the effects of controlling resemblance weights. We {find} that when $\alpha$ {is} much greater than $\beta$, region selection fits boundary resemblance well. However, it {is} easy to blur boundaries, {which is} against users' intentions. When $\alpha$ is much less than $\beta$, it {is} hard to choose color regions continuously. To gain a balance of both two resemblances, we set $\alpha = 0.3$ and $\beta = 0.7$ {after testing them in our experiments} in our proposed tool. In addition, we {evaluate} our tool under different sketch sizes. As shown in Table \ref{tab:performance}, our proposed region selection algorithm, along with the proposed {smudge} tool, {demonstrate} impressive performance in terms of frame rates. Specifically, when operating on images of size $512 \times 512$, our region selection algorithm {achieves} a remarkable frame rate of 720 frames per second (fps), while the corresponding tool {achieves} a rate of 48 fps. Similarly, when working with larger images of size $1024 \times 1024$, our algorithm {maintains} a high frame rate of 240 fps, while the {corresponding} tool {operates} at 15 fps.

\subsection{User Study}
To further investigate the effectiveness, expressiveness, and usability of the proposed algorithm and the proposed system, we conducted a user study on 12 participants. Our user study included two studies (Study 1 and Study 2). In Study 1, we compared the proposed \sysName~tool with the basic traditional tool. In Study 2, we allowed participants to paint with our system freely. The order of {paintings in both studies} was counterbalanced by a Latin square design to minimize possible learning effects. A 20-minute tutorial session was offered for each study.

\subsubsection{Study 1: Reproducing Paintings}

{\textbf{Participants and Apparatus.}}
{We recruited 12 participants (7 novices and 5 professionals) in Study 1}. Our prototype was implemented {in Python} and successively tested {on} a Surface Book 3 (supporting multi-touch gestures) with an Intel i7-1065G7 CPU and an NVIDIA GeForce GTX 1060 GPU. The input devices for painting were a Wacom digital tablet with the corresponding pen and a Microsoft Surface Book 3 with a Surface Slim pen. Since our tool supports different types of input devices, we allowed users to exploit their preferred input devices.

\textbf{Task.} We provided users with two smudge tools from the proposed system: the \sysName~tool  (SS for short) owning a dynamic brush with the same parameter setting as Sec. \ref{sec:alg} and the basic traditional smudge tool (BS for short) whose brush size {(ranging from 0 to 200 in both tools)} was decided by users. Given {a reference {painting} and a flat-filled painting obtained by decomposing the corresponding reference painting,} users were required to smudge colors in the flat-filled regions according to the reference paintings as much as they can with the two smudge tools. We provided two reference paintings with different levels of painting difficulty: an apple and an avocado (Fig. \ref{fig:study1}). The latter was harder since it included more colors and thus required more {color-smudging} operations.

\textbf{Performance Measures.} Our system recorded the following information for quantitative analysis: the completion time of each painting work, the number of color smudge operations, the number of undo operations, the number of all operations except changing brush sizes, and SUS scores. The SUS questionnaire is included in the supplemental material. In addition, a semi-structured interview was conducted to collect our study participants' subjective feelings and comments for qualitative analysis.

\textbf{Results and Discussion{.}} 
Fig. \ref{fig:study1} shows the paintings created by our study participants using the two compared smudge tools in Study 1. We adopted a paired t-test ({$p < .05$}) to analyze the statistical significance of the above-mentioned metrics in Study 1 for pairwise comparisons. 

As shown in Table \ref{tab:appleanalysistask1}, we observed that our proposed \sysName tool led to significantly less completion time. (BS: 9.42 mins vs SS: 7.34 mins, \emph{p = .0005}). This {was attributed} to the dynamic brush of our proposed algorithm, which helped the study participants to save {time} for selecting brush sizes. On average, there were significantly fewer color smudge operations using SS than BS (SS: 79.00 times vs BS: 263.83 times, \emph{p = .0006}). This result was confirmed by the participant feedback as well as from the semi-structured interview. For example, P2 commented: ``Dynamic size-adaptive brushes blended more color regions in larger regions and less in smaller regions automatically. It reduced time cost and brought about more natural shading effects.'' This followed our observation since it is challenging for non-professional participants to {select appropriate} brush parameters, causing high time cost and human labor. In addition, in terms of the number of all types of operations apart from adjusting brush size operations (only in BS), SS required significantly fewer operations than BS when painting (SS: 94.16 times vs BS: 287.54 times, \emph{p = .0004}). This was somewhat expected since dynamic size-adaptive brushes from {the \sysName tool} set the sizes automatically and generated more natural shading effects. {This was} more efficient than the process where users set brush sizes first and then used different brushes in color smudging.

\begin{table}[t]
\centering
\caption{Subjective results in {Study 1}.} 
\setlength\tabcolsep{2.5pt}
\begin{tabular}{p{1.2in}ccc}
\hline
\textbf{Metrics} &\textbf{\textbf{BS}} & \textbf{SS} & \textbf{Paired T-test}  \\ \hline
Time (min)  & 9.42 & 7.34 & p = .0005 \\
Smudge (count) & 263.83 & 79.00 &  p = .0006\\
Undo (count) & 17.33 & 11.33 & p = .1772 \\ 
All operations (count) & 287.54 & 94.16 & p = .0004  \\
\hline
\end{tabular}
\label{tab:appleanalysistask1}
\end{table}

As shown in Fig. \ref{fig:study1}, the fixed small or large brush failed to generate smooth and natural shading effects, and the large brush also blurred boundaries against our study participants' intentions. It can be seen from this figure that the painted apples with SS had {a} smoother transition from highlights to shadows and fewer artifacts in {the} boundaries of regions than the apples created with BS. As for the usability factors SS was judged as easier to use (SS: 3.70 vs BS: 2.66, \emph{p = .0081}), less complex (SS: 2.16 BS: 3.50, \emph{p = .0154}), and less inconsistent (SS: 2.08 vs BS: 3.16, \emph{p = .0352}) than BS. The study participants would like to exploit SS more frequently (SS: 3.91 vs BS: 2.66, \emph{p = .0171}).

\subsubsection{{Study 2: Freely Painting}}
{\textbf{Participants, Apparatus, and Task.}}
 In {Study 2, we recruited another 3 participants {(P1 to P3; P1 is a professional and P2-P3 are amateurs)}
 in this study. The apparatus in this study was the same as that in Study 1. The task was the free creation of paintings with our prototype system, which integrated drawing, color selection, our \sysName~tool, basic traditional smudge tool, and auxiliary functions like undoing, moving canvas, rescaling, etc.} We provided both {two smudge tools}, i.e., our \sysName~tool and basic traditional smudge tool, for users to select freely during painting to collect their preferences {on them.}  {After completing the paintings, we provided a 30-minute semi-structured interview to collect users' feedback.}

\begin{figure}[ht]
\centering
 \includegraphics[width=0.98\linewidth]{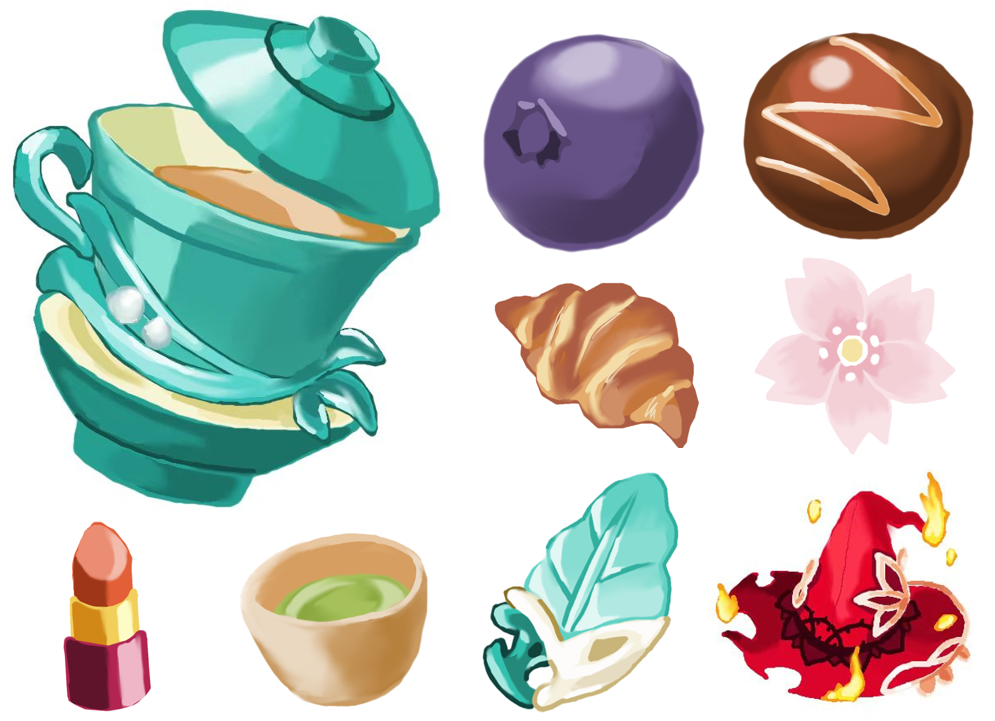}
\caption{A gallery of paintings created by the participants in Study 2. Users conducted over 90\% of the color smudge operations with the proposed \sysName~tool. }
\label{fig:showcase}
\end{figure}

{\textbf{Results and Discussion.}}
Lastly, Fig. \ref{fig:showcase} gives a gallery of digital paintings created by the novices and professionals by exploiting {the proposed prototype system} in {Study 2}. The study participants spoke highly of expressiveness, intuitiveness, {robustness}, and efficiency of our proposed tool. P1 commented: ``The \sysName~tool is awesome. It offered edge reservation and helped to recover edges {(Fig. \ref{fig:edgerecovery})} during color smudging. It saved lots of time and reduced smudging errors, so I could concentrate on painting all the time.'' Over 90\% of smudge operations were completed via the proposed \sysName~tool. We observed that the users preferred to use the basic traditional smudge tool {for} messy color regions with multiple color patches, possibly because the clustered color regions from the painting after smudging more and more times became increasingly complex and messy. In such cases, the region selection results might be slightly different from users' intentions. {P2 commented, ``I would like to exploit the \sysName~tool to smudge in most smudge operations and then use the traditional smudge tool to refine the details of some extremely tiny parts, especially a tiny part of thin and long regions.''} For step-by-step results, please see the attached demo video.

\section{Conclusion and Future Work}
In this paper, we first {proposed} the region-aware smudge algorithm for shading effects in digital painting and implemented the prototype system based on the algorithm. We tackled the challenges related to edge reservation and natural shading effects generation. Our evaluation confirmed that the proposed algorithm {allowed} the study participants to create paintings with shading effects more efficiently and intuitively in a single-layer canvas than a traditional color smudge tool. In addition, the generalized system of digital painting empowered users to paint creatively and efficiently. Last, our algorithm might not only inspire digital painting but also empower region-aware robotic painting. For example, painting robots could utilize our region-aware information in conjunction with varying controller pressures to simulate the effect of realistic brush pressure or decompose paintings into distinct areas for parallel painting tasks.
\begin{figure}[t]
\centering
\includegraphics[width=0.95\linewidth]{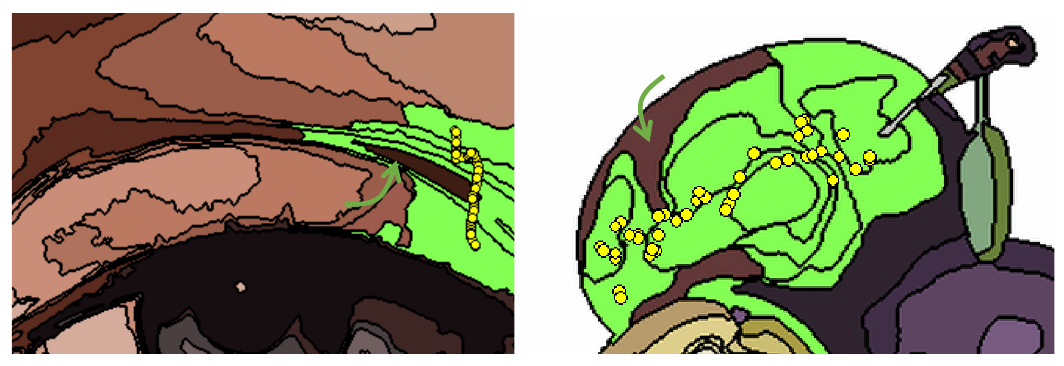}
\caption{Our proposed algorithm {selects the regions in green but fails to select the region pointed by a green arrow given the smudge path indicated by yellow circles}.
}
\label{fig:limit}
\end{figure}

Although all the study participants appreciated the efficiency and effectiveness of our proposed color-smudge tool, it remains to be improved. First, our algorithm sometimes might not cover desired regions when {smudging in thin and long} flat-filled regions {(Fig. \ref{fig:limit})}. {Under such circumstances, the traditional smudge tool might be better to smudge {those} flat-filled regions.} As Fig. \ref{fig:limit} shows, our tool failed {to select} the region pointed by a green arrow, since, considering the region resemblance and boundary resemblance put forward in our algorithm, the region only covered a little of region and edge with the partial stroke. However, {from} a global perspective, it might be better to select it. It is valuable to explore how to integrate global factors
with the proposed algorithm in further research. Second, it is better to adjust the size of the partial stroke to select color regions according to their shapes and sizes. Though the width $w$ is set to a fixed value in our current implementation, it could be a variable decided by users. A broader smudge stroke indicates a user's desire to select more regions, and the second term in Equation \ref{eq1} $\frac{\mathcal{A}(s_{t})\cap\mathcal{A}(\hat{\mathcal{T}}_{t}^*)}{\mathcal{A}(s_{t})}$ in our algorithm is designed to accommodate this by selecting more regions. Third, although our tool supports a dynamic size-adaptive brush to offer smooth shading effects on flat-filled paintings, the generated shading effects might be impacted by the shapes of flat-filled color regions. It is valuable to explore how to offer real-time shape-adaptive brushes based on different shapes of flat-filled color regions. In addition, adopting shape-adaptive brushes will be more efficient, which saves the time cost of selecting brush shapes. Last, our region selection is based on heuristic rules {and exploited fixed parameters to control resemblance weights}, and sometimes might not capture the user's intentions accurately. To gain a deeper understanding of users' intentions, we hope to {integrate machine learning techniques based on color patterns or users' habits with the proposed algorithm to further} investigate more on classifying and recognizing their intentions when they are painting.

\section{Acknowledgments}

We thank the anonymous reviews for their valuable comments. The research is partially supported by the Research Postgraduate Studentship funded by the Research Grant Council of the Hong Kong SAR Governmentand. Pengfei Xu was also supported by NSFC (62072316), NSF of Guangdong Province (2023A1515011297). 
\ifCLASSOPTIONcaptionsoff
  \newpage
\fi



%


\bibliographystyle{IEEEtran}
\bibliography{smudge}
%
\begin{IEEEbiography}[{\includegraphics[width=1in,height=1.25in,clip,keepaspectratio]{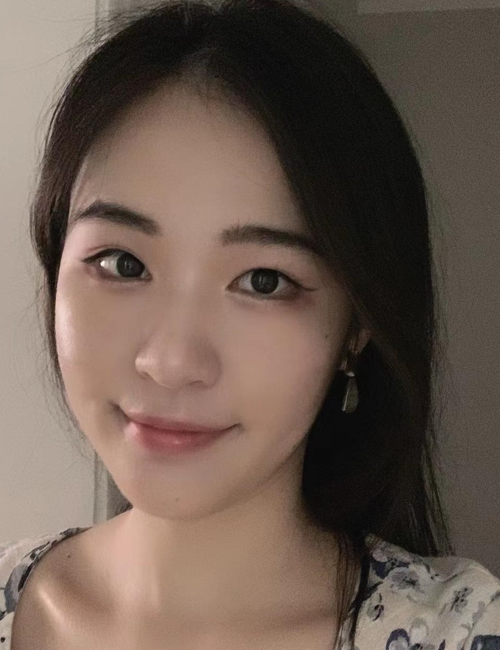}}] {Ying Jiang} 
is a Ph.D Candidate at the University of Hong Kong. Her main research interests and previous research experience include Computer Graphics (2D \& 3D sketching, modeling) and Human-Computer Interaction. 
\end{IEEEbiography}

\begin{IEEEbiography}[{\includegraphics[width=1in,height=1.25in,clip,keepaspectratio]{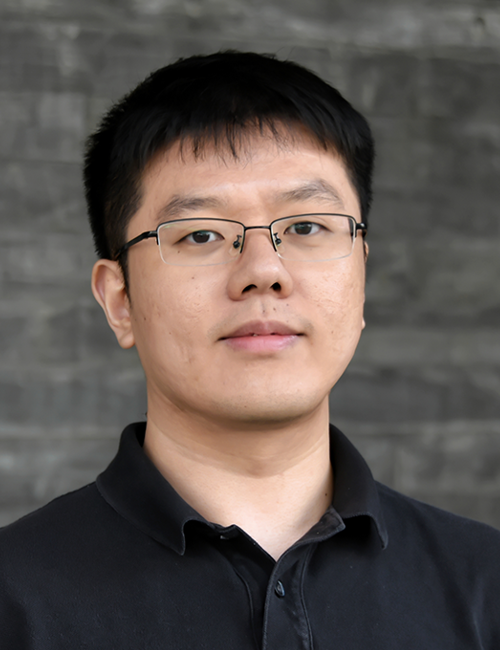}}]{Pengfei Xu} is an Associate Professor of the College of Computer Science and Software Engineering at Shenzhen University. He received his Bachelor's degree in Math from Zhejiang University, China, in 2009 and his Ph.D. degree in Computer Science from the Hong Kong University of Science and Technology in 2015. His primary research lies in Human-Computer Interaction and Computer Graphics.
\end{IEEEbiography}

\begin{IEEEbiography}[{\includegraphics[width=1in,height=1.25in,clip,keepaspectratio]{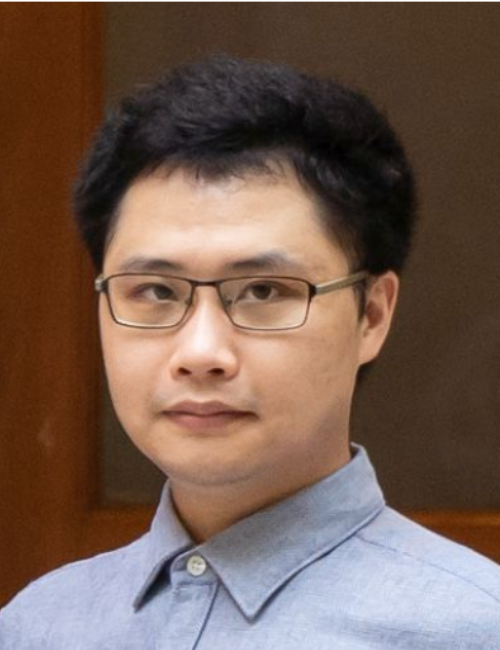}}]{Congyi Zhang}
is a postdoctoral fellow at the University of British Columbia. He received his B.Sc. degree from the School of Mathematical Science, Fudan University, in 2012, and his Ph.D. degree from the School of Electronics Engineering and Computer Science, Peking University, in 2019. His research interests include 3D reconstruction and modeling, augmented reality and virtual reality, and human-computer interaction.
\end{IEEEbiography}

\begin{IEEEbiography}[{\includegraphics[width=1in,height=1.25in,clip,keepaspectratio]{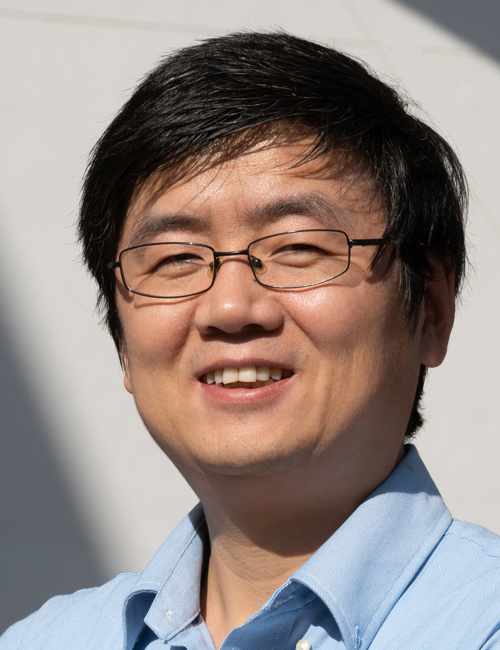}}]{Hongbo Fu}
received a BS degree in information sciences from Peking University, China, in 2002 and a PhD degree in computer science from the Hong Kong University of Science and Technology in 2007. He is a Full Professor at the School of Creative Media, City University of Hong Kong. His primary research interests fall in the fields of computer graphics and human-computer interaction. He has served as an Associate Editor of The Visual Computer, Computers \& Graphics, and Computer Graphics Forum.
\end{IEEEbiography}

\begin{IEEEbiography}[{\includegraphics[width=1in,height=1.25in,clip,keepaspectratio]{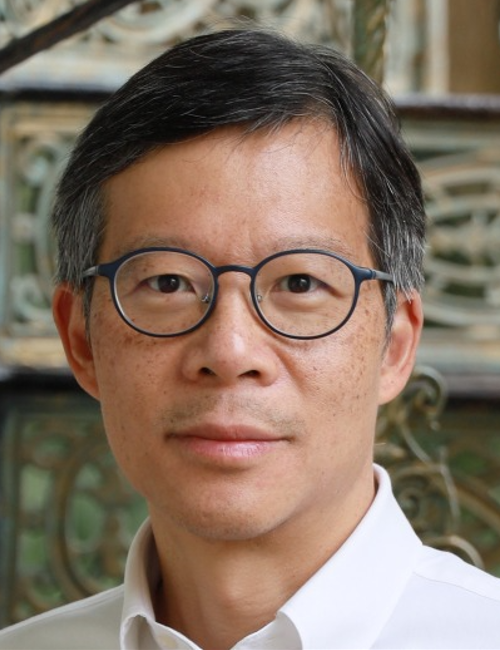}}]{Henry Lau} graduated with bachelor's degree in Engineering Science and DPhil in Robotics from the University of Oxford. He was the Associate Dean of Engineering and is currently an Honorary Associate Professor at the University of Hong Kong. His research interest includes virtual and augmented reality technology, robotics and artificial intelligence, particularly artificial immune systems (AIS).
\end{IEEEbiography}

\begin{IEEEbiography}
[{\includegraphics[width=1in,height=1.25in,clip,keepaspectratio]{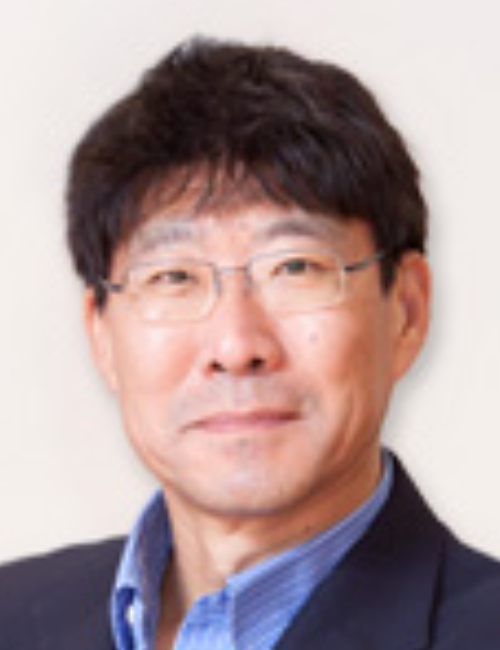}}]{Wenping Wang}
got his Ph.D. in computer science in 1992 at the University of Alberta.  His research interests include computer graphics, computer vision, geometric modeling,  robotics, and medical image processing. He has been with the University of Hong Kong from 1993 to 2020 and is now with Texas A\&M University. He is an IEEE Fellow and ACM Fellow.
\end{IEEEbiography}







\end{document}